# Optical appearance of the gravitational lens system B0218+35.7*

Frank Grundahl[1][†] and Jens Hjorth[1,2][‡]

[1] *Institute of Physics and Astronomy, University of Aarhus, DK-8000 Århus C, Denmark*
[2] *Institute of Astronomy, Madingley Road, Cambridge CB3 0HA*

**ABSTRACT**

We present high-resolution CCD images (FWHM$\approx 0''\!.6$–$0''\!.7$) of the compact gravitational lens system B0218+35.7 obtained at the Nordic Optical Telescope. Using aperture and PSF photometry we detect the two closely separated (335 mas) components A and B as well as the lensing galaxy. In the optical B seems to be very much brighter than A, opposite to the radio morphology. This suggests that most of the light from A is absorbed in the lensing galaxy, possibly by a giant molecular cloud located in the line-of-sight to component A. The lensing galaxy appears to be a late-type galaxy at a small inclination. For future variability studies we present calibrated photometry of the system and its immediate surroundings, even though the relative faintness of A will make it difficult to obtain an optical time delay even with the *Hubble Space Telescope*.

## 1 INTRODUCTION

B0218+35.7 is a gravitationally lensed flat-spectrum radio source (Patnaik et al. 1993). Its core is imaged into two components (A and B) closely separated by the record small 335 mas, and its jet gives rise to a steep-spectrum radio ring (a so-called 'Einstein ring') centred 40 mas from component B. The radio core is highly polarised and has a high Faraday rotation measure (presumably occurring in the lensing galaxy).

In the optical O'Dea et al. (1992) identified B0218+35.7 with an $r \approx 20$ object. The object was slightly resolved in 1 arcsec seeing which was interpreted as the signature of the lensing galaxy. Browne et al. (1993) found that the optical spectrum of B0218+35.7 is red, typical of a BL Lac-like object, and with narrow emission and absorption lines at a redshift of 0.6847. Neutral hydrogen (Carilli, Rupen & Yanny 1993) has been detected at $z = 0.68466$ suggesting that the lens is a gas-rich galaxy at this redshift.

Recently, several new observational results have stimulated the interest in B0218+35.7.

High-resolution (0.5 mas) VLBA observations made at 15 GHz have shown that each of the two flat-spectrum components (A,B) split up in two (A1,A2 and B1,B2) with identical structures (Patnaik, Porcas & Browne 1995). These data show that (i) A1 and A2 are tangentially distorted in the same direction, (ii) the flux ratios A1/B1$\approx$A2/B2$\approx$A/B$\approx$3.6, (iii) surface-brightness is conserved in the resolved components A2 and B2, and (iv) parity inversion is detected. These new results leave no doubt that B0218+35.7 is a textbook example of gravitational lensing. An important implication of these observations is that the lens potential is non-spherical (Patnaik et al. 1995).

Wiklind & Combes (1995) have detected molecular gas as CO, HCO$^+$, and HCN absorption at the redshift of the lens. The molecular absorption has a high filling factor of about 0.8 or larger. Furthermore, because of the small velocity spread of their spectra, Wiklind & Combes (1995) suggested that the lensing galaxy has a small inclination.

Finally, the radio-emission is known to be variable and a preliminary polarisation time delay of about 12 days has been reported (Wilkinson 1995). This short time delay and the relatively simple lensing configuration has made B0218+35.7 a prime candidate for a cosmological determination of the Hubble constant from gravitational lensing (Refsdal 1964). The redshift of the source is at present unknown.

In this Letter we present new high-resolution optical images of the system to help answer some of the questions spawned by these new developments. In particular we discuss the nature of the lensing galaxy and present evidence that A is significantly absorbed at optical wavelengths, possibly due to the passage of the light from A through a giant molecular cloud in the lensing galaxy.

---

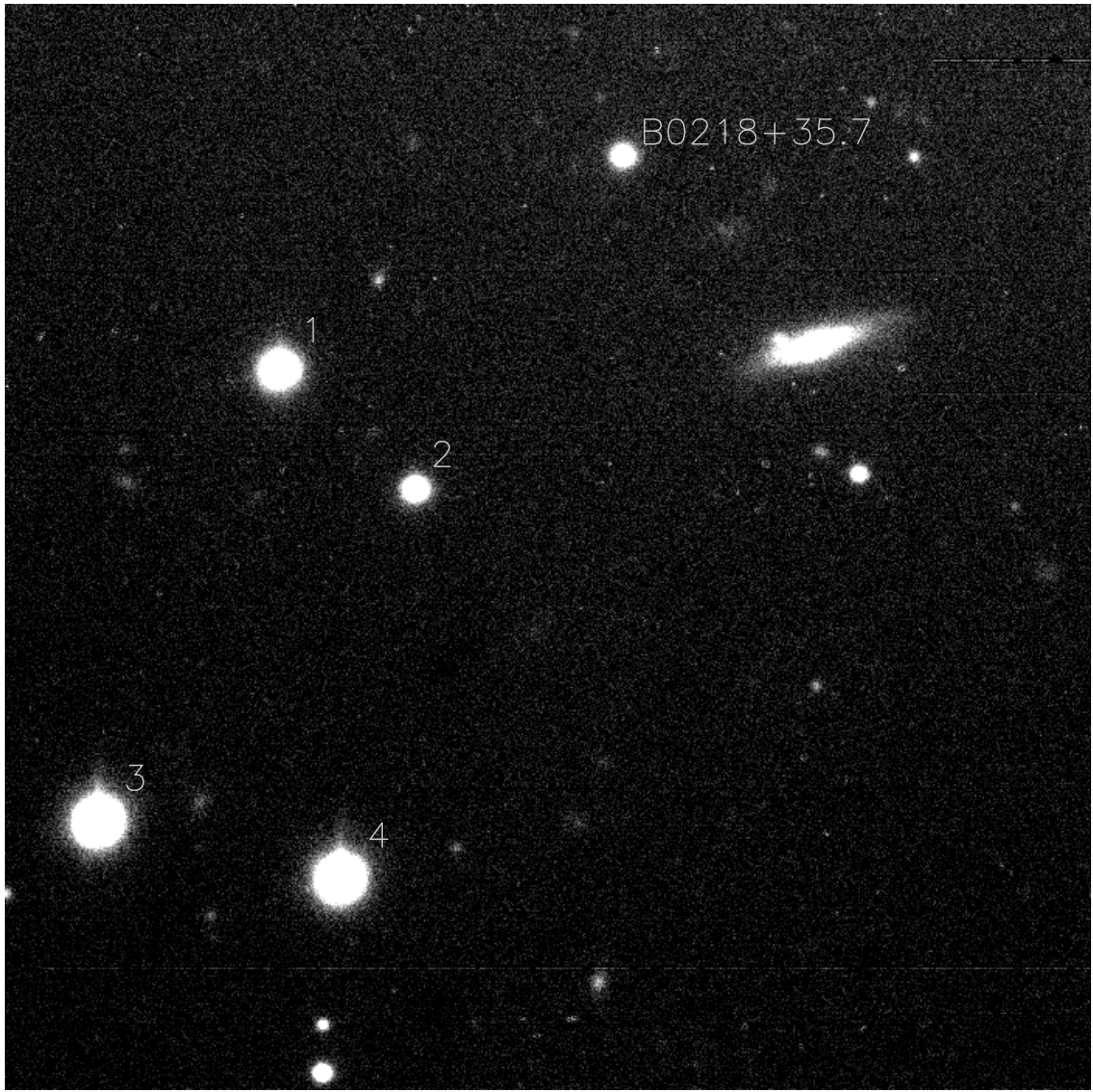

**Figure 1.** Picture of the field surrounding B0218+35.7. The section of the combined image shown here measures 700 × 700 pixels (77 arcsec × 77 arcsec) and includes the 4 stars used for defining the point spread function. North is up and east is to the left.

## 2  OBSERVATIONS AND IMAGE ANALYSIS

### 2.1  Observations

The observations were obtained at the 2.56-m Nordic Optical Telescope (NOT), La Palma, in August 1993. For the imaging we used a 2048 × 2048 pixel CCD camera built at the Copenhagen University Observatory (Brorfelde). It employed a thick, UV–coated CCD chip from Loral. We measured the gain and read-out noise to be 1.3 $e^-$ $ADU^{-1}$ and 7.7 $e^-$ (rms), respectively. The pixel size was 0.11 arcsec $pixel^{-1}$ and the total field measured 3.7 arcmin × 3.7 arcmin.

The observing log is shown in Table 1. The observations from August 14, were plagued by a very high content of Sahara dust over the observatory, causing an extinction somewhat higher than normal. Therefore no standard stars were measured on this night. On the night of August 15, the weather improved significantly, with photometric conditions and no dust visible over the observatory. The seeing this night was also slightly better than the previous night. Standard stars from the list of Landolt (1992) were observed. A total of 9 fields were measured, some of them several times. We also observed a standard field in the globular cluster M92.

### 2.2  Reductions

The images were bias subtracted and flat fielded using the average of the twilight sky flats obtained during the run. A coadded image constructed from all five frames is presented in Fig. 1. The observations were transformed to the standard Johnson/Cousins $VI$ system system using ≈ 65 measured stars from the list of Landolt (1992) and from a field in M92 (Christian et al. 1985; Davis 1994). The formal errors in the transformation are small (the standard deviation is $\sigma = 0.014$ mag), but we consider it likely that an error of 0.04 mag is more realistic. This is due to a flat field problem caused by scattered light reflected to the CCD via the inside of the primary mirror baffle at the time of observations (Grundahl & Sørensen 1995). This scattered light affected the structure of the flat fields which was seen to depend on the position of the instrument rotator (NOT is alt-az mounted).

**Table 1.** Observing log

| Date | UTC | Filter | Exposure time | FWHM |
|---|---|---|---|---|
| Aug 14 1993 | 05:07 | *I* | 600 s | 0″.68 |
| Aug 14 1993 | 05:20 | *I* | 900 s | 0″.68 |
| Aug 15 1993 | 04:35 | *V* | 1200 s | 0″.62 |
| Aug 15 1993 | 04:58 | *V* | 1200 s | 0″.62 |
| Aug 15 1993 | 05:22 | *I* | 900 s | 0″.61 |

### 2.3 Aperture photometry

Photometry of the images was carried out using DAOPHOT II and ALLSTAR (Stetson 1987). All calibrated magnitudes reported in this Letter were obtained by aperture photometry. Apertures between 2 and 44 pixels were used. In Fig. 2 'aperture growth curves', i.e., the accumulated light at a given radius relative to that at 2 pixels, for five sources in the first *V* image obtained on August 15 are shown. The aperture growth curves for the other images are of similar quality. The growth curves match very well for the 4 brightest stars in the field, and the variation in the outer parts ($R > 40$ pixels) between each successive aperture is $\approx 0.002$ mag. This shows that the background estimation is reliable.

It is immediately apparent that the aperture growth curve of B0218+35.7 does not resemble that of a single point source, but contains an extra, more extended, contribution to the light. We also note that the growth curve does not flatten very well in the outer parts. The most likely origin of this problem is that fainter objects are located within the synthetic apertures. Indeed it is possible to see 3 faint objects within 5 arcsec of B0218+35.7 (cf. Fig. 1). We have therefore estimated the magnitude of B0218+35.7 by extrapolating its aperture growth curve from a radius of 20 pixels and outward using the mean aperture growth curve for the 4 bright stars in the field. This is a reasonable approach since it is expected that the light from the putative underlying galaxy at $z = 0.68$ has fallen to a negligible level at a radius of 20 pixels (2.2 arcsec). This assertion is corroborated by the PSF photometry described below.

The photometry of B0218+35.7 and the bright stars is summarised in Table 2. We find that B0218+35.7 is 0.27 (*V*) and 0.25 (*I*) magnitudes brighter than expected if it consisted of a single point source only. This extra signal is, as we shall argue, most likely due to the lensing galaxy and the obscured A component.

### 2.4 Point spread function photometry

The four stars labelled in Fig. 1 were used as PSF stars. The magnitudes derived from point-source fitting were very close to those obtained by aperture photometry; aperture corrections were of the order of 0.02 to 0.04 mag for all sources and images discussed. This is the case for the stars as well as for the point source of B0218+35.7. Thus, this entirely different way of determining the magnitudes strongly supports the results reported in Table 2.

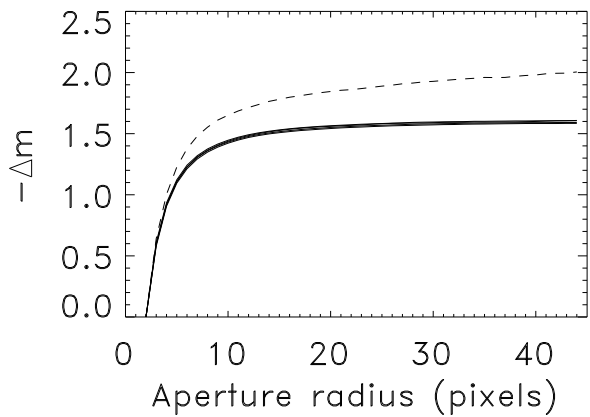

**Figure 2.** This figure shows the 'growth curves' (accumulated light as a function of distance from the centre of the source) for the first *V* image. The solid curves are for the four stars (cf. Fig. 1 and Table 2) and the dashed curve is for B0218+35.7.

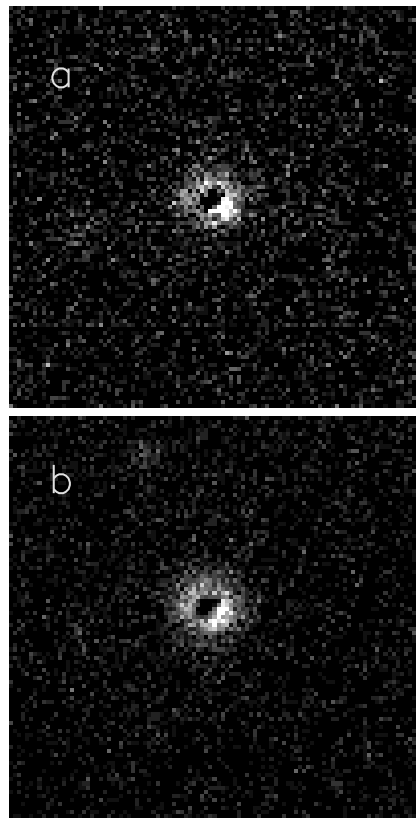

**Figure 3.** Residuals after PSF photometry of the (a) *V* and (b) *I* images. The excerpts shown here measure 100 × 100 pixels (11 arcsec × 11 arcsec). North is up and east is to the left.

Subtraction of the point source fitted to the peak of B0218+35.7 yields some conspicuous residuals. These are shown in Fig. 3 for the combined *V* and *I* images, respectively. The independent residuals in the two pass bands are very similar: There is a slight over-subtraction of the bright

**Table 2.** Photometry of B0218+35.7 and the nearby PSF stars

|              | $V$   | $I$   | $V - I$ | $\sigma$ (mag) |
|--------------|-------|-------|---------|----------------|
| Total light       | 20.35 | 18.54 | 1.81    | 0.07 |
| Point source (B)  | 20.62 | 18.79 | 1.83    | 0.07 |
| Fuzz (galaxy +A)  | 22.0  | 20.3  | 1.7     | 0.2  |
| Star 1            | 17.24 | 16.46 | 0.78    | 0.05 |
| Star 2            | 18.91 | 17.86 | 1.05    | 0.05 |
| Star 3            | 16.15 | 15.43 | 0.72    | 0.05 |
| Star 4            | 16.32 | 15.55 | 0.77    | 0.05 |

The uncertainty in the photometry is dominated by flat fielding errors ($\approx$ 0.04 mag) for the brightest objects and flux level for the faint objects.

core, some faint emission surrounding it, and an extra source of light to the south west within the extent of the fuzz. No significant residuals were visible at the positions of the subtracted (much brighter) stars which means that the residuals shown in Fig. 3 are not an artifact of the photometry. An exception is the characteristic oversubtraction ($\sim$5 percent of the peak) which arises because the background level has been underestimated. Artificially increasing the background estimate to the level of the fuzz, with subsequent fitting and subtraction of a point source makes the hole go away without affecting any of the other conclusions drawn from the PSF photometry.

## 3 DISCUSSION

### 3.1 Interpretation

We now present our interpretation of the residuals shown in Fig. 3. Some alternative interpretations are briefly discussed in Section 3.5.

The bright point source subtracted is interpreted as one of the two components A or B and the diffuse light surrounding it is interpreted as the lensing galaxy. Lensing models (e.g., Patnaik et al. 1993; Refsdal & Surdej 1994) require component B to be very close to the centre of the deflecting body. As the bright point source is located close to the centre of the diffuse light (the galaxy) it appears that B is the bright image. This assignment is supported by the fact that a significant excess of light is detected in both the $V$ and $I$ images at a position angle (roughly south west) and distance ($\approx$ 0.35 arcsec) from the core that matches the position of A relative to B known from the radio maps (Patnaik et al. 1993).

In this interpretation we have thus detected the two components A and B as well as the lensing galaxy in our optical images. B is the brighter image located close (in projection) to the center of the galaxy and A is the faint source (roughly 4 to 8 times fainter) 0.35 arcsec to the south-west of B. The galaxy has a relatively circular appearance with a diameter of 2 arcsec corresponding to 10 kpc ($H_0 = 75$ km s$^{-1}$ Mpc$^{-1}$ and $q_0 = 0.5$ is used throughout this Letter). Unfortunately, there are no bright stars in our images that can be used for astrometry to confirm our interpretation.

### 3.2 Absorption by a molecular cloud

Below we briefly discuss the possibility that A is covered by a molecular cloud in the lensing galaxy.

There is a monotonic increase in the intensity ratio A/B from 2.6 at 1.6 GHz to 3.7 at 15 GHz (see Patnaik et al. 1993; O'Dea et al. 1992; Shaffer 1994). This variation probably arises because the emitting regions at different frequencies come from different places or because of scattering by ionized gas in the lensing galaxy (or both). In any case, it is likely that the 'true' core intensity ratio is detected at high frequencies and is close to 4.

The intrinsic extent of A (A1+A2) is about 1 mas at high frequencies (Patnaik et al. 1995). This corresponds to about 5 pc at the redshift of the galaxy, i.e., it is indeed possible for a molecular cloud to cover the entire image A.

Under the two simple hypotheses that (i) A/B=4 and that (ii) A is completely absorbed whereas B is not, the filling (covering) factor should be 0.8. This is remarkably consistent with the results of Wiklind & Combes (1995) who found a lower limit to the filling factor of about 0.85. Wiklind & Combes (1995) had problems with this idea because they assumed A/B $\approx$ 3 and because they took the low frequency VLBI size (5 mas) as an indicator of the extent of A (Patnaik et al. 1993).

### 3.3 Nature of the lensing galaxy

The lensing object is probably the centre of a late-type galaxy since it is gas-rich, has a high rotation measure, and has a small mass (because of the small image separation). It is also likely that the lens has an elliptical potential because of the (i) presence of the steep spectrum Einstein ring, (ii) from VLBA observations it has been found that the lens potential in non-spherical (Patnaik et al. 1995) and (iii) the simple lens model of Narasimha & Patnaik (1993) (see Patnaik et al. 1994) requires it to be eccentric. Thus, the observations seem to require the lens to be a low-mass, gas-rich, elliptical object.

An edge-on spiral satisfies these requirements, but Wiklind & Combes (1995) argued that the spiral had to be at a low inclination (i.e., face on). Our data (see Fig. 3) suggest that the galaxy is indeed more or less circular (in projection) which seems to support the assertion of Wiklind & Combes (1995). Assuming that half the light of the fuzz in Fig. 3 comes from component A with $V - I = 1.8$ then the galaxy has $I = 21.1$ and $V - I = 1.6$. Following Glazebrook et al. (1994) we can convert the observed magnitudes to a rest absolute magnitude of $M_B \approx -20.2$ and a rest colour of $U - B \approx 0.0$, which are typical values for a luminous spiral galaxy. As the lensing constraints on the mass and the elliptical potential are only valid within the Einstein radius, i.e., within $0\rlap{.}''17 \approx 800$ pc, the eccentric potential would then have to be formed by the bulge of the spiral or possibly a bar. This seems to be a possible explanation as only a total mass of a few times $10^{10}$ M$_\odot$ is required to reproduce the observed images. This possibility could be tested in the upcoming *Hubble Space Telescope* (*HST*) observations.

### 3.4 Prospects for determining the Hubble constant

B0218+35.7 could be ideal for determination of the Hubble constant. In order to determine $H_0$ a time delay and the full lensing configuration must be known.

Unfortunately, A is so faint that even regular *HST* observations may be unsuccessful in determining the time delay. However, radio polarization observations will probably pin down the value of the time delay (Wilkinson 1995).

Concerning the source redshift the optical spectrum did not provide a convincing redshift determination (Browne et al. 1993). Therefore, an IR spectrum seems to be needed to determine this important quantity. The implication of our proposed scenario with a molecular cloud in front of A is that A should emerge from behind the cloud in the IR and provide a factor of 4 more photons to help achieve an IR spectrum of sufficient signal-to-noise ratio. This also opens up the possibility of determining an IR time-delay using adaptive optics.

The lens potential appears to be well constrained and so B0218+35.7 remains a very important cosmological object and a prime candidate for a cosmological determination of the Hubble constant.

### 3.5 Alternative interpretations of the optical images

We finally discuss some possible alternative interpretations of the images presented in this Letter.

The bright point source could be the core of the lensing galaxy. The arguments against this possibility is that the optical spectrum is not that of a galaxy core (Browne et al. 1993; Stickel & Kühr 1993) and that such a bright system at $z = 0.68$ does not appear compatible with a low-mass, gas-rich system.

Alternatively, the bright point source is component A. In this case one would have to explain why A is located near the center of the galaxy, and one would have to account for the missing light at the position of B as well as the extra light on the opposite side of B.

The possibility that the fuzz surrounding the point source is not the galaxy but the Einstein ring can be safely dismissed because of the high spectral index of the ring and its small diameter of only 335 mas compared to $\approx 2$ arcsec for the fuzz.

The possibility that the entire system can in fact be accounted for by two point sources with an intensity ratio of 4:1 (i.e., that our photometry is somehow misleading) has been excluded by reductions of simulated data.

## 4 CONCLUSION

We have imaged the gravitational lens system B0218+35.7 at high angular resolution using the Nordic Optical Telescope. We have detected a bright point source (interpreted as B) near the center of the lensing galaxy which appears to be an almost circular object (in projection). It is likely to be a face-on late-type galaxy. The A component is also detected but is found to be very much fainter than B. We have proposed that this is due to absorption by a molecular cloud in the lensing galaxy. This explanation is consistent with the millimetre spectra of Wiklind & Combes (1995). The photometry is summarised in Table 2.

These results and interpretations have several implications and predictions: (i) A should emerge behind the cloud in IR images making the object 4 times brighter than expected from its visual magnitude. This could help determine the source redshift by obtaining an IR spectrum. (ii) The fact that lensing requires an eccentric lens potential whereas the system seems to be fairly symmetric indicates that the lensing galaxy could be a spiral with a bulge or a bar. This can be tested in future *HST* images. (iii) Higher S/N millimetre data should establish to what extent the spectra of Wiklind & Combes (1995) are saturated. This has implications for the value of the filling factor, which in the present simple scenario (in which B is not significantly absorbed) should not be much higher than 0.8. (iiii) Astrometry of the field should also establish whether the bright point source is indeed component B, as suggested in this Letter.

When these questions have been settled and when the time delay has been securely determined, the prospects for a cosmological determination of the Hubble constant from B0218+35.7 appear very promising.

## ACKNOWLEDGMENTS

We thank Ian Browne, Alok Patnaik and Tommy Wiklind for very fruitful discussions and exchange of results prior to publication. FG acknowledges financial support from the Danish Board for Astronomical Research. JH was supported by the Danish Natural Science Research Council (SNF).

## REFERENCES

Browne I. W. A., Patnaik A. R., Walsh D., Wilkinson P. N., 1993, MNRAS, 263, L32
Carilli C. L., Rupen M. P., Yanny B., 1993, ApJ, 412, L59
Christian C. A., Adams M., Barnes J. V., Butcher H., Hayes D. S., Mould J. R., Siegel M., 1985, PASP, 97, 363
Davis L., 1994, private communication
Glazebrook K., Lehár J., Ellis R., Aragón-Salamanca A., Griffiths R., 1994, MNRAS, 270, L63
Grundahl F., Sørensen A. N., 1995, in preparation
Landolt A. U., 1992, AJ, 104, 340
Narasimha D., Patnaik A. R., 1993, in Surdej, J. et al., eds, Proc. 31st Liège Int. Astroph. Coll., Université de Liège, Liège, p. 295
O'Dea C. P., Baum S. A., Stanghellini C., Dey A., van Breugel W., Deustua S., Smith E. P., 1992, AJ, 104, 1320
Patnaik A. R., Browne I. W. A., King L. J., Muxlow T. W. B., Walsh D., Wilkinson P., 1993, MNRAS, 261, 435
Patnaik A. R., Porcas R. W., Browne I. W. A., Muxlow T. W. B., Narasimha D., 1994, in Zensus J. A., Kellermann K. I., eds, Compact Extragalactic Radio Sources, Socorro, NM, USA, p. 129
Patnaik A. R., Porcas R. W., Browne I. W. A., 1995, MNRAS, 274, L5
Refsdal S., 1964, MNRAS, 128, 307
Refsdal S., Surdej J. 1994, Rep. Prog. Phys., 57, 117
Shaffer D. B., 1994, in Zensus J. A., Kellermann, K. I., eds, Compact Extragalactic Radio Sources, Socorro, NM, USA, p. 132
Stickel M., Kühr H., 1993, A&AS, 101, 521
Stetson P. B., 1987, PASP, 99, 191
Wiklind T., Combes F., 1995, A&A, in press
Wilkinson P. N., 1995, in Highlights of Astronomy, Vol. 10, in press